# Automatic Segmentation of the Kidneys and Cystic Renal Lesions on Non-Contrast CT Using a Convolutional Neural Network

Lucas Aronson, BS[1], Ruben Ngnitewe Massa'a, MD[1], Syed Jamal Safdar Gardezi, PhD[1], Andrew L. Wentland, MD, PhD[1,2,3]

[1]Department of Radiology, University of Wisconsin School of Medicine & Public Health, Madison, WI, USA

[2]Department of Medical Physics, University of Wisconsin School of Medicine & Public Health, Madison, WI, USA

[3]Department of Biomedical Engineering, University of Wisconsin School of Medicine & Public Health, Madison, WI, USA

Corresponding Author
Andrew L. Wentland, MD, PhD
Department of Radiology
600 Highland Avenue
University of Wisconsin School of Medicine & Public Health
Madison, WI 53792
Email: alwentland@wisc.edu
Phone: 608-265-2021

**Abstract**

**Objective**: Automated segmentation tools are useful for calculating kidney volumes rapidly and accurately. Furthermore, these tools have the power to facilitate large-scale image-based artificial intelligence projects by generating input labels, such as for image registration algorithms. Prior automated segmentation models have largely ignored non-contrast computed tomography (CT) imaging. This work aims to implement and train a deep learning (DL) model to segment the kidneys and cystic renal lesions (CRLs) from non-contrast CT scans.

**Methods**: Manual segmentation of the kidneys and CRLs was performed on 150 non-contrast abdominal CT scans. The data were divided into an 80/20 train/test split and a deep learning (DL) model was trained to segment the kidneys and CRLs. Various scoring metrics were used to assess model performance, including the Dice Similarity Coefficient (DSC), Jaccard Index (JI), and absolute and percent error kidney volume and lesion volume. Bland-Altman (B-A) analysis was performed to compare manual versus DL-based kidney volumes.

**Results**: The DL model achieved a median kidney DSC of 0.934, median CRL DSC of 0.711, and total median study DSC of 0.823. Average volume errors were 0.9% for renal parenchyma, 37.0% for CRLs, and 2.2% overall. B-A analysis demonstrated that DL-based volumes tended to be greater than manual volumes, with a mean bias of +3.0 ml (±2 SD of ±50.2 ml).

**Conclusion**: A deep learning model trained to segment kidneys and cystic renal lesions on non-contrast CT examinations was able to provide highly accurate segmentations, with a median kidney Dice Similarity Coefficient of 0.934.

**Keywords**: deep learning; kidney segmentation; artificial intelligence; convolutional neural networks.

**Introduction**

Kidney segmentation on cross-sectional imaging is of interest for volume calculations and dosimetry. Furthermore, kidney segmentations can be used as input labels for image registration algorithms; such image registration is particularly helpful for radiologists in evaluating the kidneys in multiphase CT examinations, such as CT urography. Finally, segmentations can facilitate the assessment of renal morphology and aid in localizing pathologies [1]. Manual kidney segmentation is a time-consuming process [2] and is prone to intra- and interobserver variability. Rapid, automated methods of segmentation are much needed.

Prior efforts have been made to automatically segment the kidneys and cystic renal lesions (CRLs) from CT imaging with deep learning (DL) models [3], [4], [5], [6]. One such model, termed the Medical Imaging Segmentation with Convolutional Neural Networks (MIScnn), yielded a median Dice coefficient of 0.9544 for the kidneys and 0.7912 for kidney tumors [7]. However, the MIScnn model, as well as other previously published models, have focused on segmentation via solely contrast-enhanced CT images. A model that can provide reliable segmentation of the kidneys and kidney lesions on non-contrast CT imaging would be beneficial.

The purpose of this study was to train a deep learning model to provide automatic segmentations of kidneys and CRLs on non-contrast CT data sets.

**Materials and Methods**

*Patient Cohort*

This Health Insurance Portability and Accountability Act–compliant study was approved by our institutional review board. The need for written informed consent was waived given the retrospective nature of the study.

Non-contrast CT datasets curated for this study were derived from a clinical database of screening CT colonography examinations between 2000 and 2021—acquired as previously described [8], [9]. Briefly, non-contrast CT images were acquired with a multidetector CT scanner (GE Healthcare, Waukesha, WI, USA) at 120 kVp with a variable tube current (mA) and with a mean effective dose of 5 mSv. Axial supine CT slices were reconstructed to 3.75-5 mm in thickness. Clinical reports (per abdominal fellowship-trained radiologists) from the CT scans were screened to identify two cohorts: cases with CRLs and cases without CRLs. For each cohort, studies were selected randomly. In total, 150 studies were selected, consisting of 35.3% cases without CRLs and 64.7% cases with CRLs (Figure 1).

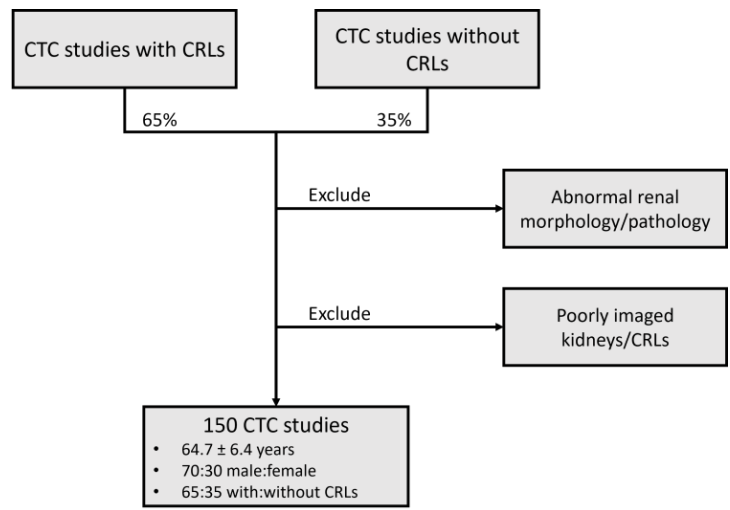

**Figure 1.** Studies were randomly selected from two large cohorts of non-contrast CT colonography studies. Approximately 65% of cases came from one cohort of patients whose kidneys contained cystic renal lesions. The remaining 35% of studies came from another cohort

of patients who did not have any cystic renal lesions. Studies with abnormal renal morphology, renal pathology, or which contained poorly or incompletely imaged kidneys or renal lesions were excluded. Thus, 150 studies were used for model development, with a mean age of 64.7 years, a roughly 70:30 male:female ratio, and a roughly 65:35 with:without cystic renal lesion ratio.

*Data Preparation*

Ground truth segmentations of the kidneys and CRLs were performed in ITK-SNAP. The active contour feature of ITK-SNAP was used to first generate approximate segmentations of kidneys and CRLs based on initial user inputs. These initial segmentations were then manually adjusted to delineate the structures fully and accurately. For the kidney parenchyma segmentation, the structures of the renal hilum (vasculature and collecting system) were excluded manually. These segmentations were initially performed by a student (LA) and subsequently adjusted as needed by a fellowship-trained abdominal radiologist with five years of experience in abdominal imaging (AW). Furthermore, the number and location of cystic renal lesions was initially performed by a student (LA) and subsequently adjusted as needed via the same fellowship-trained abdominal radiologist (AW).

The segmentations were parsed into three classes, with lesions assigned values of 2, kidney parenchyma assigned a value of 1, and background assigned a value of 0. The segmentation masks and CT image sets were saved as compressed Neuroimaging Informatics Technology Initiative (NIfTI) files.

*Deep Learning Model*

The MIScnn package was selected as the base DL framework [3], [6], [7], [10]. Via this framework, a standard 3D U-Net architecture [11] was selected. Studies were resampled to uniform voxel dimensions (1.62 x 1.62 mm in-plane resolution and 3.22 mm slice thickness), as previously described [12]. Voxel attenuation values were normalized using a z-score prior to additional preprocessing and clipped to the range (-79, 304 Hounsfield units) in order to reduce computational burden; values outside of this Hounsfield unit range were considered irrelevant for the sake of kidney and renal lesion segmentation. Two cycles of data augmentation were performed to increase sample size and improve model robustness. Eight unique data augmentation techniques were implemented per cycle, including scaling (between 0.85 and 1.25 times the original image size), rotation (between -15 and +15 degrees), elastic deformation (stretching/squishing), mirroring, brightness, contrast, gamma (adjusts both brightness and contrast), and Gaussian noise [7].

*Training the Model*

Patient-level data were divided via a random 80/20 train/test split (Figure 2A). As such, the algorithm was trained on 120 studies using a 300 epoch three-fold cross-validation and Tversky loss function [13] to tune hyperparameters. As training progressed, the best-performing model based on Tversky loss was continuously saved. Training was run on an in-house NVIDIA (Santa Clara, CA, USA) Tesla V100 SXM2 GPU with 32 GB of RAM.

*Volumetric Calculations*

Volumes of the kidneys and CRLs were calculated for both manual and model-based segmentations by first counting the number of voxels in the segmentation masks corresponding to kidneys (intensity = 1) or lesions (intensity = 2). Then, voxel dimensions were extracted from the NIfTI header files. Finally, the number of kidney and lesion voxels were multiplied by the voxel volume to derive the volumes of kidneys and CRLs automatically (Figure 2B).

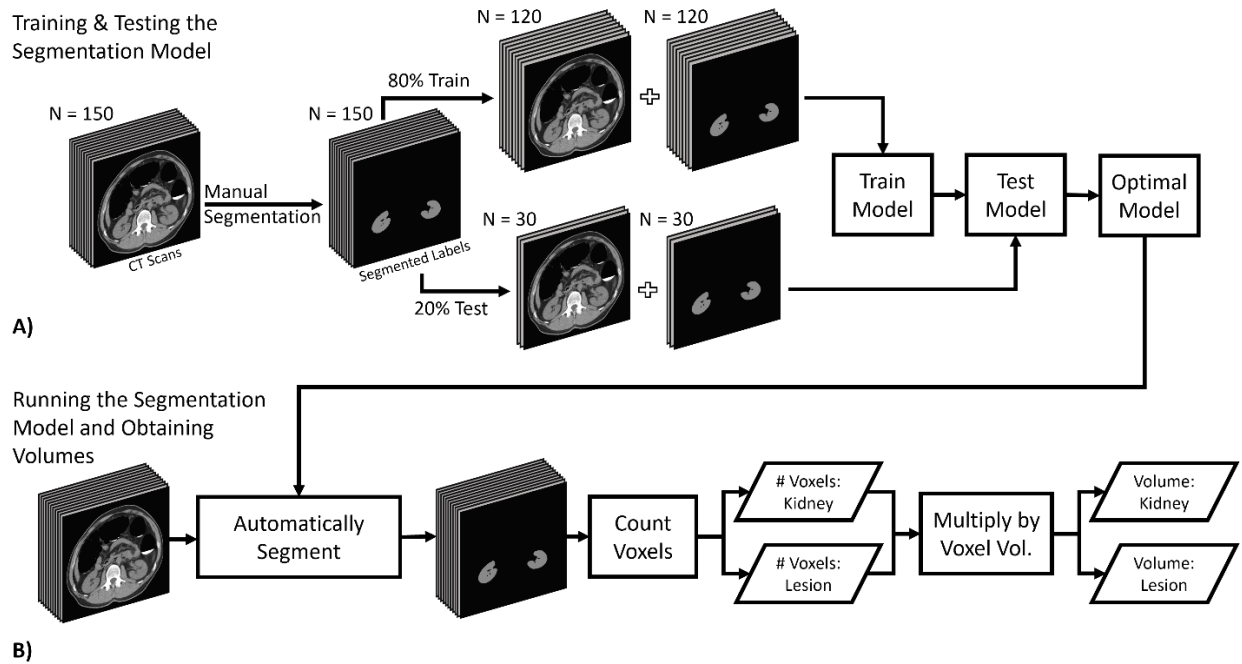

**Figure 2**. (**A**) Kidneys and renal lesions were manually segmented within a cohort of 150 non-contrast CT scans. These studies were then randomly split such that 80% of CT scans and segmentations were used for training the deep learning (DL) model and 20% were reserved for testing. The DL model was then trained and tested on the respective cohorts to optimize a segmentation model. (**B**) Kidney and renal lesion volumes were calculated automatically by first counting the number of voxels that comprise each segmentation class and then by multiplying the number of voxels by the voxel volume.

*Evaluating the Model*

The 30 test studies were passed through the trained algorithm to generate DL-based predicted segmentations and volumes. The DL-based segmentations were then evaluated based on the Dice Similarity Coefficient (DSC) and Jaccard Index (JI), which were summarized using boxplots. For the identification of CRLs, the binary question of whether a lesion was identified or missed was tabulated and compared to the reference standard of CRL identification by a fellowship-trained abdominal radiologist. A confusion matrix was produced to summarize true positive (TP), true negative (TN), false positive (FP), and false negative (FN) detection rates for CRLs. These values were used to calculate accuracy, sensitivity, specificity, positive predictive value (PPV), and negative predictive value (NPV) of CRL detection. The number of right versus left CRLs, as well as the number of endophytic versus exophytic CRLs, were tabulated. An exophytic CRL was defined as having >50% of the cyst extending beyond the renal capsule.

DL-based volume predictions for kidneys and renal lesions were evaluated via both absolute and percent error using Bland-Altman (B-A) analysis [14]. These errors were calculated using DL-based segmentation volume subtracted from manual segmentation volume such that negative values correspond to situations where the DL-based method overestimated the manual method. A t-test was then performed to test the null hypothesis that the DL-based volumes were not significantly different from the manual volumes, with a p-value of significance set at 0.05.

**Results**

The cohort of 150 non-contrast CT studies used in this study consisted of 69.8% male and 30.2% female patients with a mean ± standard deviation age of 64.7 ± 6.4 years, ranging from 54 to 73 years. Of this whole cohort, there were 91 (46.0%) and 107 (54.0%) CRLs in the right and left kidneys, respectively (Table 1). Among these lesions, 58 (29.3%) and 140 (70.7%) were endophytic and exophytic, respectively (Table 1).

**Table 1.** Number and percent of cystic renal lesions in right and left kidneys among the whole dataset (training set + testing set). Number and percent of endophytic (Endo.) and exophytic (Exo.) cystic renal lesions among the whole dataset.

|         | Training Set |       |       | Testing Set |       |       | Total |       |       |
|---------|-------|-------|-------|-------|-------|-------|-------|-------|-------|
| Class   | Right | Left  | Total | Right | Left  | Total | Right | Left  | Total |
| Lesions | 75    | 78    | 153   | 16    | 29    | 45    | 91    | 107   | 198   |
|         | 49.0% | 51.0% |       | 35.6% | 64.4% |       | 46.0% | 54.0% |       |
| Class   | Endo. | Exo.  | Total | Endo. | Exo.  | Total | Endo. | Exo.  | Total |
| Lesions | 40    | 113   | 153   | 18    | 27    | 45    | 58    | 140   | 198   |
|         | 26.1% | 73.9% |       | 40.0% | 60.0% |       | 29.3% | 70.7% |       |

Results from the testing set yielded mean kidney DSC of $0.931 \pm 0.013$ (median = 0.934), mean CRL DSC of $0.584 \pm 0.292$ (median = 0.711), and total mean study DSC of $0.758 \pm 0.147$ (median = 0.823) (calculated as the mean of the average of the kidney and CRL segmentation for each study across all test studies). Mean JIs for the kidneys, CRLs, and total were $0.872 \pm 0.023$, $0.467 \pm 0.268$, and $0.669 \pm 0.148$, respectively (Table 2, Figure 3). Cases without CRLs yielded a mean total DSC of $0.929 \pm 0.013$ (median = 0.929), and cases with CRLs yielded mean total DSC of $0.758 \pm 0.153$ (median = 0.823). Similarly, cases without CRLs yielded mean total JIs of $0.868 \pm 0.023$ (median = 0.868), and cases with CRLs yielded mean total JIs of $0.670 \pm 0.146$ (median = 0.715).

**Table 2.** Medians and standard deviations of Dice similarity coefficients (DSC) and Jaccard indices (JI) for a deep learning segmentation model of kidneys with and without lesions on non-contrast CT data sets. Mean ± standard deviation (median).

|  | Kidneys without Lesions | | Kidneys with Lesions | | All Kidneys | |
|---|---|---|---|---|---|---|
|  | DSC | JI | DSC | JI | DSC | JI |
| Kidney | 0.929 ± 0.013 (0.929) | 0.868 ± 0.023 (0.868) | 0.932 ± 0.013 (0.935) | 0.873 ± 0.023 (0.879) | 0.931 ± 0.013 (0.934) | 0.872 ± 0.023 (0.876) |
| Lesion | N/A | N/A | 0.584 ± 0.292 (0.711) | 0.467 ± 0.268 (0.551) | 0.584 ± 0.292 (0.711) | 0.467 ± 0.268 (0.551) |
| Total | 0.929 ± 0.013 (0.929) | 0.868 ± 0.023 (0.868) | 0.758 ± 0.153 (0.823) | 0.670 ± 0.146 (0.715) | 0.758 ± 0.147 (0.883) | 0.669 ± 0.148 (0.793) |

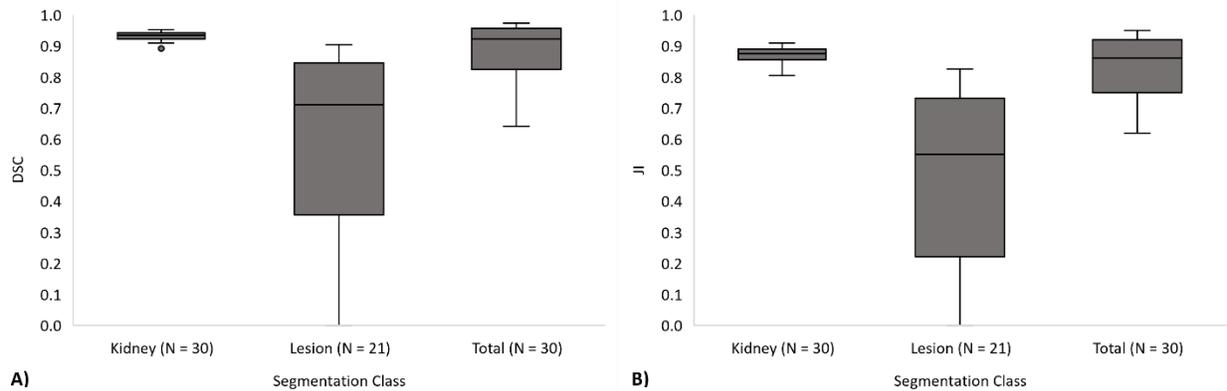

**Figure 3.** Boxplots of **(A)** Dice Similarity Coefficients (DSC) and **(B)** Jaccard Indices (JI) for automatic segmentation of kidneys and renal lesions on non-contrast CT data sets using a deep learning model. N represents the number of cases that apply to each category.

CRL detection was measured by counting lesions that were at least partially segmented by the algorithm as lesions. Within the test set of 30 subjects, there were 30 CRLs. The segmentation model correctly identified 23 CRLs and correctly did not identify any CRLs in eight subjects. However, the model missed 7 CRLs and inaccurately identified 7 CRLs in which none was present. Metrics for CRL detection were calculated based on the confusion matrix (Table 3), yielding an accuracy of 0.69, sensitivity of 0.77, specificity of 0.53, PPV of 0.77, and NPV of 0.53.

**Table 3.** Confusion matrix for automatic renal lesion detection via a deep learning segmentation model on non-contrast CT data sets. True positive (TP), true negative (TN), false positive (FP), and false negative (FN).

|  |  | Prediction | |
|---|---|---|---|
|  |  | Positive | Negative |
| Ground Truth | Positive | TP = 23 | FN = 7 |
| Ground Truth | Negative | FP = 7 | TN = 8 |

Percent error volume measurements for each class were as follows: kidneys = -0.94%, CRLs = -37.01%, and total = -2.17% (Table 4). Overall, DL-based volumes were generally overestimated compared to manual volumes. B-A analysis [14] comparing manual versus DL-based kidney volumes demonstrated that DL-based volumes tended to be greater than manual volumes, with a mean bias of -3.0 ml (±2 SD of ±50.2 ml) (Figure 4A). B-A analysis of manual versus DL-based CRL segmentation volumes also showed that DL-based CRL volumes tended to be greater than manual CRL volumes, with a mean bias of -4.2 ml (±2 SD of ±41.3 ml) (Figure 4B). Percent error B-A analysis reflected these tendencies for DL-based segmentations to overestimate manual segmentations in both kidney volumes (mean bias of -0.9% with ±2 SD of ±15.8%, Figure 4C) and lesion volumes (mean bias of -22.9% with ±2 SD of ±190.6%, Figure 4D). Several example cases comparing DL-based segmentations and manual segmentations are displayed in Figure 5.

**Table 4.** Means and standard deviations of kidney and lesion volumes as derived from manual segmentations versus segmentations derived from a deep learning (DL) model. Mean and standard deviation of volumetric percent errors calculated as 100 * (Ground Truth - Predicted) / Ground Truth.

| Class | Manually Derived Volume (ml) | DL-Based Volume (ml) | % Error |
|---|---|---|---|
| Kidney | 324.6 ± 75.9 | 327.6 ± 80.4 | -0.9 ± 5.9 |
| Lesion | 11.5 ± 14.2 | 15.7 ± 23.6 | -37.0 ± 66.8 |
| Total | 336.0 ± 81.9 | 343.3 ± 86.2 | -2.2 ± 5.3 |

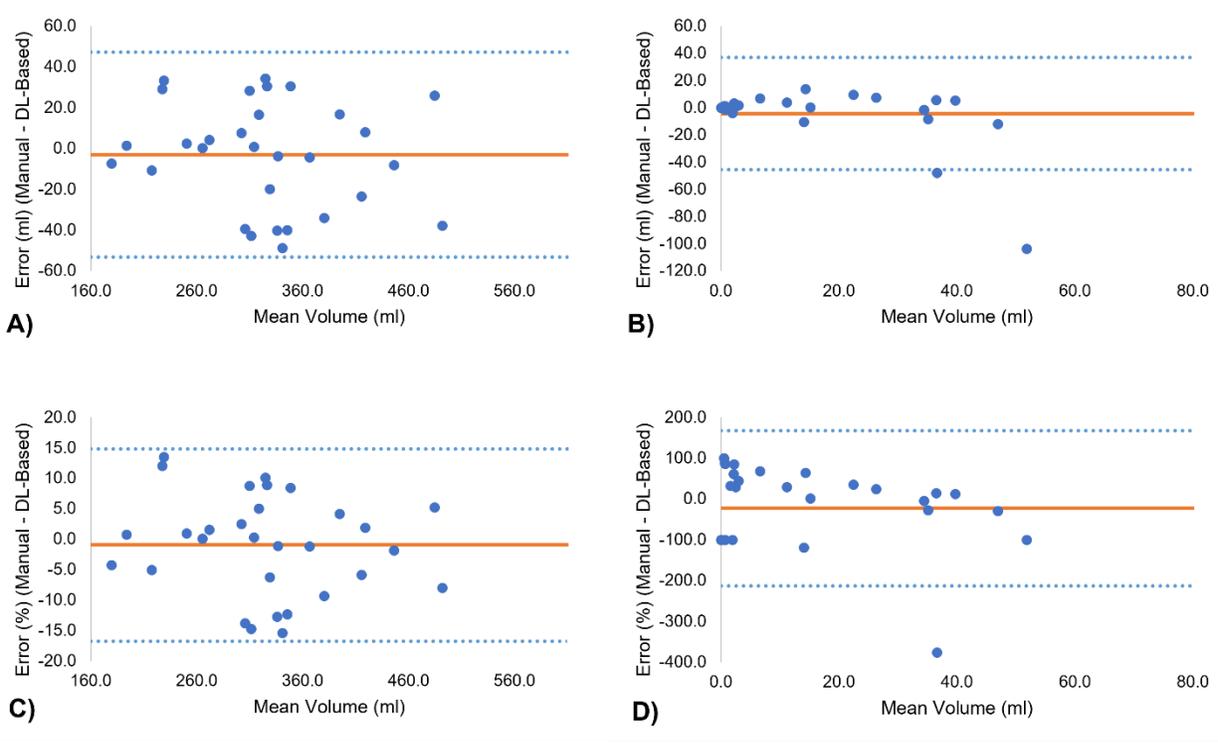

**Figure 4.** Bland-Altman plots assessing agreement of kidney and renal lesion volumes derived from manual segmentations versus deep learning (DL)-based segmentations. (**A**) and (**B**) depict absolute volumes for kidneys and lesions, respectively, and (**C**) and (**D**) depict percent volumes for kidneys and lesions, respectively.

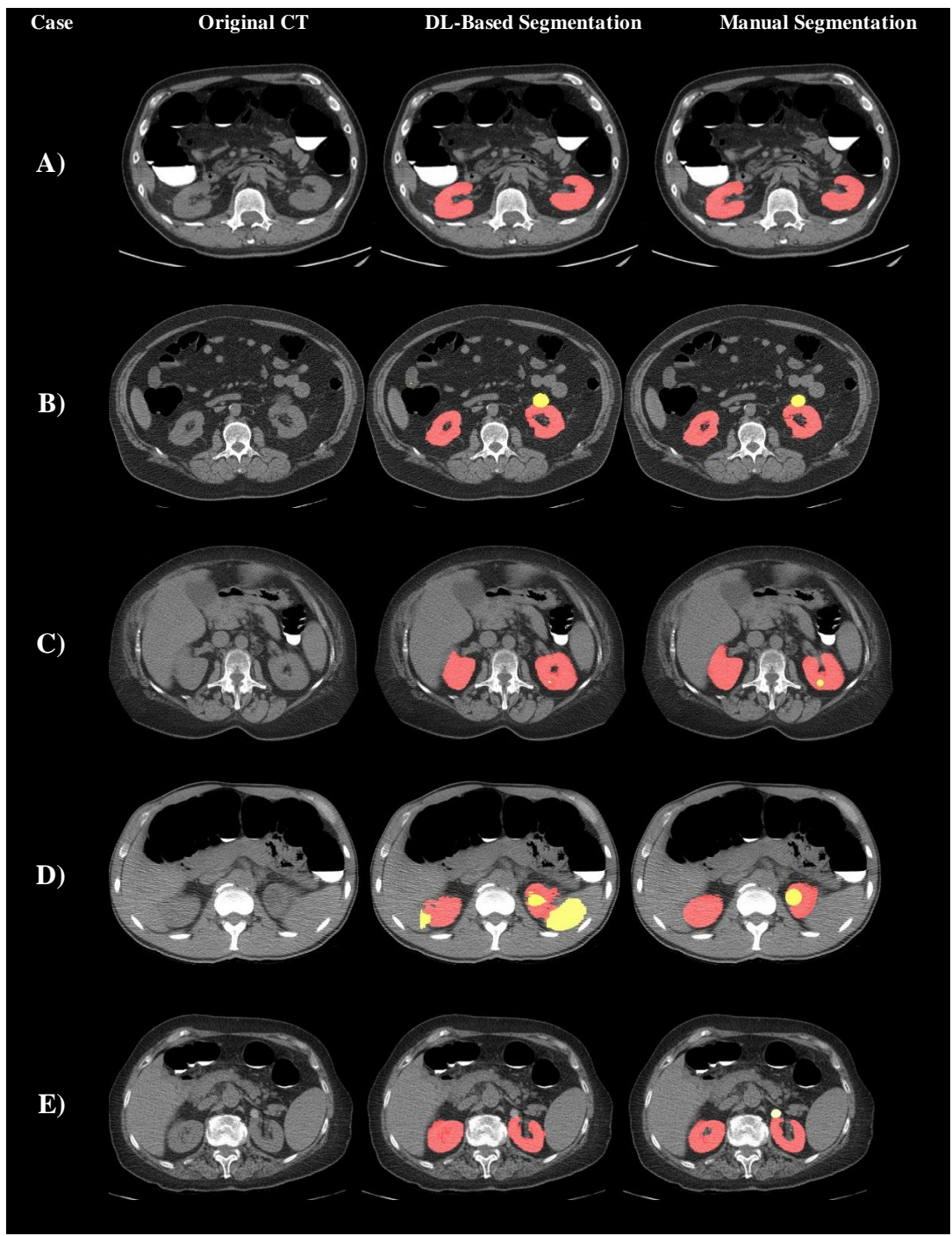

**Figure 5.** Example cases, where rows correspond to different cases and columns represent (left) original CT image, (middle) DL-Based predicted segmentations, and (right) manual segmentations. Renal parenchyma is shown in red and cystic renal lesions (CRLs) are shown in yellow. Total study Dice similarity coefficients (DSCs) are as follows: **(A)** Kidneys without any CRLs (DSC = 0.947). **(B)** One exophytic CRL (DSC = 0.923). **(C)** Poorly segmented CRL (DSC

= 0.492). **(D)** Erroneous segmentation of the liver and spleen as CRL (DSC = 0.514). **(E)** Undetected CRL, which was hyperattenuating as compared to adjacent renal parenchyma (DSC = 0.462).

**Discussion**

In this study, the MIScnn Python package [7] was modified and retrained for segmenting kidneys and cystic renal lesions on non-contrast CT data sets. The retrained model yielded quality segmentation scores for kidneys (DSC > 0.9) and lesions (DSC > 0.7).

Prior efforts in the field of kidney segmentation from CT have focused on contrast-enhanced imaging. Intravenous contrast greatly assists in segmentation due to the heightened attenuation differences between kidney and surrounding/background tissues such as perinephric fat, cystic lesions, the collecting system, and the renal sinus [15]. Kramer and Müller achieved kidney and lesion DSCs of 0.93 and 0.68, respectively, for arterial phase contrast-enhanced CTs based on the same DL framework [3], [7]. Liu et al. used liver and spleen segmentations to assist in delineating kidneys from surrounding structures on CTC studies [16], given that many automated kidney segmentation algorithms tend to fail where the liver and spleen abut the kidneys. Their work achieved a mean kidney DSC of 0.89; however, the model relied on additional segmentations of the liver and spleen as well as the assumption of good body symmetry. Finally, da Cruz et al. used a DL approach as well as scope reduction image processing techniques to segment the kidneys from arterial phase contrast-enhanced CT imaging to reduce false positive detection [4]. This team achieved a kidney DSC of 0.93 on an external dataset.

The results presented in this study mark a first-of-its-kind tool for segmenting the kidneys and CRLs from non-contrast CT, and thus could not be directly compared to any other study. Our DSCs for kidney segmentation (mean = 0.931) were comparable to other high-performing segmentation algorithms, despite the lack of IV contrast. Renal lesion segmentation was challenging both due to the lack of IV contrast (thus, less difference in attenuation between lesions and renal parenchyma) and the presence of small lesions. Although the DSCs for CRLs were not exceptionally high (mean = 0.584) and the detection accuracy was relatively low (0.689), DSCs and detection accuracy are likely to improve with more training and image processing techniques. While the segmentation of CRLs is not of clinical interest in patients without polycystic kidney disease, the ability to parse CRLs from renal parenchyma improves the overall segmentations of the kidneys themselves and the accuracy of renal volumes.

Several example cases are provided in Figure 5. DL-based segmentation of kidneys without CRLs (Figure 5A) tended to yield high segmentation scores, with mean DSCs of $0.929 \pm 0.013$. Occasionally the DL-based segmentations were inaccurate and tended to include adjacent organs (Figure 5D); it is postulated that such failures may be due to a paucity of perinephric fat. Increasing the size of the training cohort would likely mitigate such errors. While DL-based segmentations for CRLs yielded mean DSCs of $0.584 \pm 0.292$, some CRLs were incompletely segmented (Figure 5C versus 5B). This inaccuracy may relate to whether the lesions were endophytic versus exophytic (Table 1), which is an area of further investigation. Finally, our training data did not specifically include hyperattenuating or solid renal lesions; the model failed to identify an incidental hyperattenuating renal lesion in one of the test set cases (Figure 5E). The fact that this lesion was not detected by the algorithm also demonstrates the algorithm's weighting of attenuation in deciding whether to classify a lesion or not.

This study had several limitations. All subjects in the cohort were between the ages of 54 and 73 years and imaged at a single institution with a low dose CT protocol [8]. Similarly, lack of CT scanner variability limits model generalizability. Furthermore, the scan protocol uses high-density oral contrast [8], which occasionally led to beam-hardening artifact that partially obscured the kidneys and renal lesions. Further training of the DL model on non-contrast CT data sets without such high-density oral contrast may improve results. A small sample size was used in this study given the time-consuming nature of manual segmentations; data augmentation techniques were used to mitigate the impact of the small sample size. Furthermore, no validation set was used for training of hyperparameters given the relatively small sample size and thus the sizes of the training and testing sets were maximized. The absence of a validation set may lead to overfitting. The use of a validation set in the context of a larger cohort is an area of future work. Finally, most of the lesions present in this dataset were simple cystic lesions. It is likely that solid lesions with very little attenuation difference compared to renal parenchyma would not be well-segmented with the currently trained model.

**Significance**

A deep learning model was trained for segmenting kidneys and cystic renal lesions on non-contrast CT image sets. This model achieved excellent segmentations of the kidneys, with a median DSC of 0.93. Segmentation of the cystic renal lesions yielded a lower median DSC of 0.71. Despite these lower lesion DSCs, the DL model was able to detect the presence of renal lesions with high sensitivity of 0.77. This trained segmentation model also automatically calculated the volumes of both kidneys and cystic renal lesions and achieved mean volume errors of -0.94% and -37.01% for kidneys and cystic renal lesions, respectively.

**Model Access**

Raw data, including images and segmentations, are available upon request. The MIScnn DL framework is available on an open-source GitHub page [10].

**Disclosures**

The authors have no relevant disclosures.